\documentclass{ws-ijmpb}

\begin{document}

\markboth{Lucia \v{C}anov\'a {\it et al}.} {Exact results of the
mixed-spin Ising model on a decorated square lattice}

%
\catchline{}{}{}{}{}
%

\title{EXACT RESULTS OF THE MIXED-SPIN ISING MODEL ON A DECORATED SQUARE LATTICE
       WITH TWO DIFFERENT DECORATING SPINS OF INTEGER MAGNITUDES}

\author{LUCIA \v{C}ANOV\'A, JOZEF STRE\v{C}KA, and MICHAL JA\v{S}\v{C}UR}

\address{Department of Theoretical Physics and Astrophysics, Faculty of Science, \\
           P.~J.~\v{S}af\'{a}rik University, Park Angelinum 9, 040 01 Ko\v{s}ice, Slovak
           Republic\\lucia.canova@pobox.sk}

\maketitle

\begin{history}
\received{Day Month Year}
\revised{Day Month Year}
\end{history}

\begin{abstract}
The mixed-spin Ising model on a decorated square lattice with two
different decorating spins of the integer magnitudes $S_{\rm B} = 1$
and $S_{\rm C} = 2$ placed on horizontal and vertical bonds of the
lattice, respectively, is examined within an exact analytical
approach based on the generalized decoration-iteration mapping
transformation. Besides the ground-state analysis,
finite-temperature properties of the system are also investigated in
detail. The most interesting numerical result to emerge from our
study relates to a striking critical behaviour of the spontaneously
ordered 'quasi-1D' spin system. It was found that this quite
remarkable spontaneous order arises when one sub-lattice of the
decorating spins (either $S_{\rm B}$ or $S_{\rm C}$) tends towards
their 'non-magnetic' spin state $S = 0$ and the system becomes
disordered only upon further single-ion anisotropy strengthening.
The effect of single-ion anisotropy upon the temperature dependence
of the total and sub-lattice magnetization is also particularly
investigated.
\end{abstract}


\section{Introduction}
\label{intro} Over the last six decades, a considerable research
interest has been devoted to determine a critical behaviour and
other statistical properties of various lattice-statistical models,
which enable a deeper understanding of order-disorder phenomena in
magnetic solids. In this respect, the planar Ising model takes a
prominent place, since it represents perhaps the simplest
lattice-statistical model for which a complete exact solution is
known since the Onsager's pioneering work\cite{Ons44}. Exactly
soluble planar Ising models\cite{Bax82} provide a convincing
evidence for many controversial results predicted in the phase
transition theory. Besides, these models also provide a good testing
ground for many approximative theories. It should be stressed,
nevertheless, that a precise treatment of two-dimensional (2D) Ising
models is often connected with considerable difficulties, which
relate to the usage of sophisticated mathematical methods when
applying them to more complex models describing, for instance, spin
systems with interactions beyond nearest neighbours\cite{Oit81},
frustrated spin systems\cite{Lie86}, or higher-spin models with (or
without) single-ion anisotropy and biquadratic
interactions\cite{Hor86}. Up to now, exact results for Ising models
on the square\cite{Ons44}, triangular and honeycomb\cite{Hou50},
kagom\'e\cite{Syo51}, extended kagom\'e\cite{Huc86}, bathroom-tile
\cite{Uti51}, orthogonal-dimer\cite{Str05} and ruby\cite{Lin83}
lattice, as well as on various irregular 2D lattices, such as union
jack (centered square)\cite{Vak65}, pentagonal\cite{Uru02},
square-kagom\'e\cite{Sun06}, or two topologically distinct
square-hexagonal\cite{Lin88} lattices, have been obtained.

The Ising systems consisting of mixed spins of different magnitudes,
which are usually called also as mixed-spin Ising models, belong
among the most interesting extensions of the standard spin-$1/2$
Ising model. These models have recently enjoyed a great scientific
interest predominantly due to their rich critical behaviour they
display. Another aspect, which started to attract a scientific
interest to the mixed-spin Ising models, relates to a theoretical
modeling of magnetic structures suitable for describing a
ferrimagnetism of a certain class of insulating magnetic materials.
In this respect, the ferrimagnetic mixed-spin Ising models are very
interesting also from the experimental point of view.

Despite a significant amount of effort, there are known only few
examples of exactly solvable mixed-spin Ising models, yet. Using the
generalized forms of decoration-iteration and star-triangle mapping
transformations, the mixed spin-$1/2$ and spin-$S$ ($S \geq 1$)
Ising models on the honeycomb, diced and several decorated planar
lattices were exactly examined long years ago\cite{Fis59}. Later on,
both the aforementioned transformation procedures were generalized
in order to account for the single-ion anisotropy effect. In this
way generalized mapping transformations were employed to exactly
investigate the influence of uniaxial and biaxial single-ion
anisotropies on magnetic properties of the mixed-spin Ising systems
on the honeycomb\cite{Gon85}, bathroom-tile\cite{Str06} or diced
lattice\cite{Jas05} and several decorated planar lattices
\cite{Jas98,Oit05}. To the best of our knowledge, these are the only
mixed-spin planar Ising models with generally known exact solutions
with exception to several mixed-spin Ising models on the Bethe
(Cayley tree) lattices studied within an approach based on exact
recursion equations\cite{Sil91}. Among the remarkable models, for
which a precise solution is restricted to a certain subspace of
interaction parameters only, one should also mention the mixed-spin
Ising model on the union jack lattice treated within the mapping
onto a symmetric (zero-field) eight-vertex model\cite{Lip95}.

Exactly solvable mixed-spin Ising models on 2D lattices, the bonds
of which are decorated in various fashion by additional spins, are
therefore of particular research interest (see Ref.~\cite{Syo72} and
references cited therein). Among the systems, which belong to the
class of exactly solved decorated Ising models, one could mention,
at least, the original ferrimagnetic model introduced by Syozi and
Nakano\cite{Syo55}, partly\cite{Syo68} and multiply\cite{Syo66}
decorated models showing reentrant phase transitions, ANNNI models
\cite{Hus81}, diluted models of ferromagnetism\cite{Syo65},
decorated model systems with classical $\nu$-dimensional vector
spins\cite{Gon82}, Fisher's super-exchange model and its other
variants\cite{Fis60}, as well as the models with higher decorating
spins\cite{Fis59}. It is worth mentioning that exact solutions of
these spin model systems have furnished answers to questions
interesting from the academic point of view (scaling and
universality hypothesis, reentrant phase transitions) as well as
from the experimental viewpoint (dilution, technological application
of ferrimagnets). The great potential of ferrimagnets with respect
to technological applications has also stimulated exploration of the
single-ion anisotropy effect upon ferrimagnetic features of the
mixed spin-$1/2$ and spin-$S$ Ising models on the wholly
\cite{Jas98} and partly\cite{Oit05} decorated lattices.

Recently, Kaneyoshi\cite{Kan01} has proposed another notable example
of the decorated Ising model on a square lattice, the horizontal and
vertical bonds of which are occupied by decorating spins of
different magnitudes. Up to now, this model system has been examined
by the use of the approach based on the differential operator
technique\cite{Hon79}, whereas an accuracy of obtained results
determined the Bethe-Peierls-Weiss approximation used for the
undecorated lattice\cite{Kan01}. The purpose of this work is
therefore to extend the class of exactly solvable Ising models by
providing an accurate solution for the mixed-spin Ising model on the
square lattice with decorating spins of two different magnitudes
$S_{\rm B}$ and $S_{\rm C}$. Exact results for the model system
under consideration are obtained by the use of the generalized
decoration-iteration transformation\cite{Syo72} constituting an
exact correspondence with an effective spin-$1/2$ Ising model on the
anisotropic square (rectangular) lattice, the exact solution of
which is well known\cite{Ons44}. Within the framework of the used
mapping method, we will focus our attention first of all on the
influence of the single-ion anisotropy on the critical behaviour and
phase diagrams. Besides, temperature dependence of the total and
sub-lattice magnetization will be also particularly examined.

The outline of this paper is as follows. In Section \ref{model}, the
detailed description of the model system will be presented and then,
some basic aspects of the used transformation method will be also
shown. Section \ref{results} deals with the interpretation of the
most interesting numerical results for the mixed-spin model system
with decorating spins of two different integer magnitudes $S_{\rm B}
= 1$ and $S_{\rm C} = 2$. Finally, some concluding remarks are
mentioned in Section \ref{conclusion}.

\section{Model and method}
\label{model}

\begin{figure}[bt]
\centerline{\psfig{file=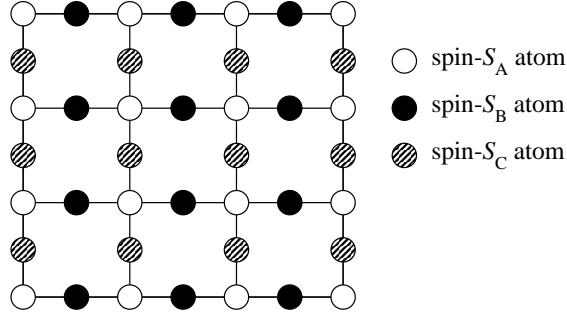,width=3.0in}} \vspace*{-0.7cm}
\caption{Part of the mixed-spin Ising model on the decorated square
lattice. The white circles denote spin-$1/2$ atoms, while the black
and hatched circles denote the decorating spin-$S_{\rm B}$ and
spin-$S_{\rm C}$ atoms, respectively.} \label{fig:1}
\end{figure}

We consider the mixed-spin Ising model on a decorated square lattice
with decorating spins of two different magnitudes $S_{\rm B}$ and
$S_{\rm C}$ placed on its horizontal and vertical bonds,
respectively, as is schematically illustrated in Fig.~\ref{fig:1}.
In this figure, the vertex sites of the original square lattice
labeled by white circles are occupied by the spin-$S_{\rm A}$ atoms,
while the decorating sites denoted by black and hatched circles are
occupied by the spin-$S_{\rm B}$ and spin-$S_{\rm C}$ atoms,
respectively. The total Hamiltonian of the model system defined upon
the underlying square lattice reads
\begin{eqnarray}
{\cal{H}} =
   J_{\rm AB}\sum_{(i,m)}^{2N} S_{i}^{z} S_{m}^{z}
 + J_{\rm AC}\sum_{(i,n)}^{2N} S_{i}^{z} S_{n}^{z} -
 D_{\rm B}\sum_{m \in {\rm B}}^{N} \left( S_{m}^{z} \right)^2
  - D_{\rm C}\sum_{n \in {\rm C}}^{N} \left( S_{n}^{z} \right)^2 \,,
\label{eq:H1}
\end{eqnarray}
where $S_{i}^{z} = \pm 1/2$,  $S_{m}^{z} = -S_{\rm B}, -S_{\rm B} +
1, \ldots, S_{\rm B}$ and $S_{n}^{z} = -S_{\rm C}, -S_{\rm C} + 1,
\ldots, S_{\rm C}$ denote three different Ising spin variables of
the spin-$S_{\rm A}$, spin-$S_{\rm B}$ and spin-$S_{\rm C}$ atoms
located at $i$th, $m$th and $n$th lattice site, respectively. The
first two summations are carried out over the nearest-neighbour A--B
and A--C pairs, respectively, while the other two summations run
over the decorating B and C lattice sites. Accordingly, the
parameters $J_{\rm AB}$ and  $J_{\rm AC}$ stand for exchange
couplings between the nearest A--B and A--C neighbours, and the
terms $D_{\rm B}$ and $D_{\rm C}$ measure a strength of the uniaxial
single-ion anisotropy acting on the spin-$S_{\rm B}$ and
spin-$S_{\rm C}$ atoms, respectively. Finally, $N$ denotes a total
number of sites of the original square lattice.

The crucial step of our calculation represents the calculation of
the partition function of the system. In view of further
manipulations, it is useful to write the total Hamiltonian
(\ref{eq:H1}) in the form
\begin{eqnarray}
{\cal{H}} &=& \sum_{m = 1}^{N} {\cal{H}}_m^{{\rm B}} + \sum_{n =
1}^{N} {\cal{H}}_n^{{\rm C}}\,, \label{eq:H2}
\end{eqnarray}
where the first (second) term represents the summation over bond
Hamiltonians each involving all interaction terms associated with
the spin-$S_{\rm B}$ ($S_{\rm C}$) atoms residing the $m$th ($n$th)
decorating site of the lattice
\begin{eqnarray}
{\cal{H}}_k^{\rm p} &=&
 J_{{\rm Ap}} S_{k}^{z}\left( S_{i}^{z} + S_{i + 1}^{z} \right) - D_{\rm p} \left( S_{k}^{z} \right)^2\,,
\label{eq:Hkp}
\end{eqnarray}
where ${\rm p} = {\rm B}$ or ${\rm C}$ and $k = m$ or $n$. The
parameters ${\rm p}$ and $k$ in Eq.~(\ref{eq:Hkp}) specify,
respectively, the decorating atom and its position in the lattice.
Therefore, it is quite evident that if $k = m$, then one considers
${\rm p} = {\rm B}$ and if $k = n$, one sets ${\rm p} = {\rm C}$.
The partition function of the considered mixed-spin system can be
partially factorized into the product
\begin{eqnarray}
{\cal{Z}} &=& \sum_{\{S_{\rm A}\}} \prod_{m,n}^{N} \sum_{S_m^z = -
S_{\rm B}}^{S_{\rm B}}\hspace{-0.2cm}\exp(-\beta{\cal{H}}_m^{\rm B})
\hspace{-0.2cm} \sum_{S_n^z = - S_{\rm C}}^{S_{\rm
C}}\hspace{-0.2cm}\exp(-\beta{\cal{H}}_n^{\rm C})\,, \label{eq:Z}
\end{eqnarray}
where $\beta = 1/(k_{\rm B}T)$, $k_{\rm B}$ is Boltzmann's constant
and $T$ stands for the absolute temperature. The product in
Eq.~(\ref{eq:Z}) is taken over all decorating lattice sites occupied
by the spin-$S_{\rm B}$ and spin-$S_{\rm C}$ atoms, while the symbol
$\sum_{\{S_{\rm A}\}}$ means the summation over all available spin
configurations of the spin-$S_{\rm A}$ atoms. Finally, the next two
summations are performed over all $(2S_{\rm B} + 1)$ and $(2S_{\rm
C} + 1)$ possible spin states of the decorating spin-$S_{\rm B}$ and
spin-$S_{\rm C}$ atoms, respectively. The structure of the latter
relation immediately implies a possibility of applying the
generalized decoration-iteration mapping transformation
\cite{Fis59,Syo72}
\begin{eqnarray}
\sum_{S_k^z = - S_{\rm p}}^{S_{\rm
p}}\hspace{-0.2cm}\exp(-\beta{\cal{H}}_k^{\rm p}) = A_{\rm
p}\exp\left(\beta R_{\rm p} S_{i}^{z}S_{i+1}^{z}\right)\,.
\label{eq:DIT}
\end{eqnarray}
From the physical point of view, the mapping (\ref{eq:DIT}) removes
all interaction parameters associated with one decorating spin
($S_{\rm B}$ or $S_{\rm C}$) and replaces them by a new effective
interaction ($R_{\rm B}$ or $R_{\rm C}$) between the remaining
vertex spins $S_{i}^{z}$ and $S_{i+1}^{z}$. Notice that a general
validity of the mapping relation (\ref{eq:DIT}) necessitates a
self-consistency condition to be satisfied, which means that it must
hold independently of the spin states of both vertex spins
$S_{i}^{z}$ and $S_{i+1}^{z}$. It can be readily proved that a
direct substitution of four possible spin configurations of vertex
spins $S_{i}^{z}$ and $S_{i+1}^{z}$ into the formula (\ref{eq:DIT})
gives just two independent equations, which unambiguously determine
the mapping parameters $A_{\rm p}$ and $R_{\rm p}$
\begin{eqnarray}
A_{\rm p} = (W_{\rm p1}W_{\rm p2})^{1/2}\,, \qquad \beta R_{\rm p} =
2\ln\left(\frac{W_{\rm p1}}{W_{\rm p2}}\right)\,, \label{eq:AR}
\end{eqnarray}
where the functions $W_{\rm p1}$ and $W_{\rm p2}$ are expressed as
\begin{eqnarray}
W_{\rm p1} &=& \sum_{n = -S_{\rm p}}^{S_{\rm p}}\hspace{-0.2cm}
\exp(\beta D_{\rm p} n^2) \cosh\left(\beta n J_{{\rm Ap}}\right)\,, \\
W_{\rm p2} &=& \sum_{n = -S_{\rm p}}^{S_{\rm p}}\hspace{-0.2cm}
\exp(\beta D_{\rm p} n^2)\,. \label{eq:V1V2}
\end{eqnarray}
After straightforward substitution of the transformation relation
(\ref{eq:DIT}) into the formula (\ref{eq:Z}), one easily obtains an
exact relation between the partition function ${\cal{Z}}$ of the
mixed-spin Ising model on the decorated square lattice and the
partition function ${\cal{Z}}_{0}$ of the undecorated spin-$1/2$
Ising model on the corresponding rectangular lattice with two
different nearest-neighbour couplings $R_{\rm B}$ and $R_{\rm C}$ in
the horizontal and vertical directions, respectively,
\begin{eqnarray}
{\cal{Z}}(\beta, J_{\rm AB}, J_{\rm AC}, D_{\rm B}, D_{\rm C}) =
A_{\rm B}^NA_{\rm C}^{N} {\cal{Z}}_{0}(\beta, R_{\rm B}, R_{\rm
C})\,. \label{eq:ZZ0}
\end{eqnarray}
The mapping relation (\ref{eq:ZZ0}) represents the central result of
our calculation, since it formally completes the exact solution of
the partition function ${\cal{Z}}$ with regard to the known exact
result for the partition function ${\cal{Z}}_{0}$ of the spin-$1/2$
Ising model on the rectangular lattice\cite{Ons44}. Actually, the
equation (\ref{eq:ZZ0}) can be utilized for calculation of some
important physical quantities, such as Gibbs free energy, internal
energy, magnetization, correlation functions, specific heat, etc.
Furthermore, by combining Eq.~(\ref{eq:ZZ0}) with commonly used
exact mapping theorems\cite{Bar88} and the differential operator
technique\cite{Hon79}, one easily proves a validity of following
exact relations for the sub-lattice magnetization $m_{\rm A}$,
$m_{\rm B}$ and $m_{\rm C}$ coincided to the spin-$S_{\rm A}$,
spin-$S_{\rm B}$, and spin-$S_{\rm C}$ atom of the mixed-spin Ising
model, respectively:
\begin{eqnarray}
m_{\rm A} &\equiv& \langle S_i^z \rangle = \langle S_i^z \rangle_0 =
m_{\rm 0} \,,
\label{eq:mA}\\
m_{\rm B} &\equiv& \langle S_m^z \rangle =  2 m_{\rm A} F_{\rm
B}(J_{\rm AB}) \,,
\label{eq:mB}\\
m_{\rm C} &\equiv& \langle S_n^z \rangle =  2 m_{\rm A} F_{\rm
C}(J_{\rm AC}) \,. \label{eq:mC}
\end{eqnarray}
Above, the symbols $\langle \ldots \rangle$ and $\langle \ldots
\rangle_0$ denote standard canonical averages performed over the
mixed-spin Ising model defined by the Hamiltonian (\ref{eq:H1}) and
its corresponding spin-$1/2$ Ising model on the rectangular lattice,
respectively. The functions $F_{\rm p}(x)$ (${\rm p} = {\rm A}, {\rm
B}$) stand for
\begin{eqnarray}
F_{\rm p}(x) = -\frac{\sum_{n = -S_{\rm p}}^{S_{\rm p}} n\exp(\beta
D_{\rm p} n^2) \sinh (\beta n x)} {\sum_{n = -S_{\rm p}}^{S_{\rm p}}
\exp(\beta D_{\rm p} n^2) \cosh (\beta n x)} \,. \label{eq:Fp}
\end{eqnarray}
It should be pointed out that with regard to Eq.~(\ref{eq:mA}), the
sub-lattice magnetization $m_{\rm A}$ directly equals to the
corresponding spontaneous magnetization $m_0$ of the spin-$1/2$
Ising model on the rectangular lattice with the effective horizontal
and vertical coupling constants given by Eq.~(\ref{eq:AR}). The
exact result for the magnetization of this model system is known for
several years\cite{Pot52}. In consequence of that, exact solutions
of both the sub-lattice magnetization $m_{\rm B}$ and $m_{\rm C}$
are formally completed.

Finally, we briefly mention an analytical condition determinating a
critical behaviour of the considered mixed-spin system. It is quite
obvious from the explicit expression of
Eqs.~(\ref{eq:mA})--(\ref{eq:mC}) that all sub-lattice magnetization
tend necessarily to zero if and only if the sub-lattice
magnetization $m_{\rm A}$ goes to zero. Accordingly, the critical
temperature can be located from the condition which is consistent
with the known Onsager's exact solution for the critical temperature
of the spin-$1/2$ Ising model on the rectangular lattice
\cite{Ons44}
\begin{eqnarray}
\sinh(\beta_{\rm c}R_{\rm B}/2)\sinh(\beta_{\rm c}R_{\rm C}/2) =
1\,, \label{eq:Tc}
\end{eqnarray}
where $\beta_{\rm c} = 1/(k_{\rm B}T_{\rm c})$ is the inverse
critical temperature and $T_{\rm c}$ denotes the critical
temperature of the considered model system.

\section{Results and discussion}
\label{results}

Before proceeding to a discussion of the most interesting results it
is worth mentioning that the results derived in the preceding
section hold regardless of whether ferromagnetic or
antiferromagnetic interactions are assumed, irrespective of values
of the decorating spins $S_{\rm B}$ and $S_{\rm C}$, as well as even
if both the single-ion anisotropy parameters $D_{\rm B}$ and $D_{\rm
C}$ are taken independently of each other. By imposing $S_{\rm B} =
S_{\rm C}$, $J_{\rm AB} = J_{\rm AC}$ and $D_{\rm B} = D_{\rm C}$,
our results reduce to those acquired for the mixed-spin Ising model
on symmetrically decorated square lattices\cite{Jas98}. Therefore,
we will restrict ourselves in the present paper only to the
particular case with the decorating spins of two different integer
magnitudes, more specifically, to the case when $S_{\rm B} = 1$ and
$S_{\rm C} = 2$. Moreover, in what follows we will restrict both the
exchange parameters $J_{\rm AB}$ and $J_{\rm AC}$ to positive
values. To simplify further discussion, we will also reduce the
number of free parameters entering in the Hamiltonian (\ref{eq:H1})
by imposing the same single-ion anisotropy parameter acting on both
kinds of decorating spins ($D_{\rm B} = D_{\rm C} = D$). Other
particular cases, in which both the decorating spins are supposed to
be half-odd-integer, or integer and half-odd-integer, respectively,
will be explored in separate works\cite{Can06,Str06b}.

\begin{figure}[bt]
\centerline{\psfig{file=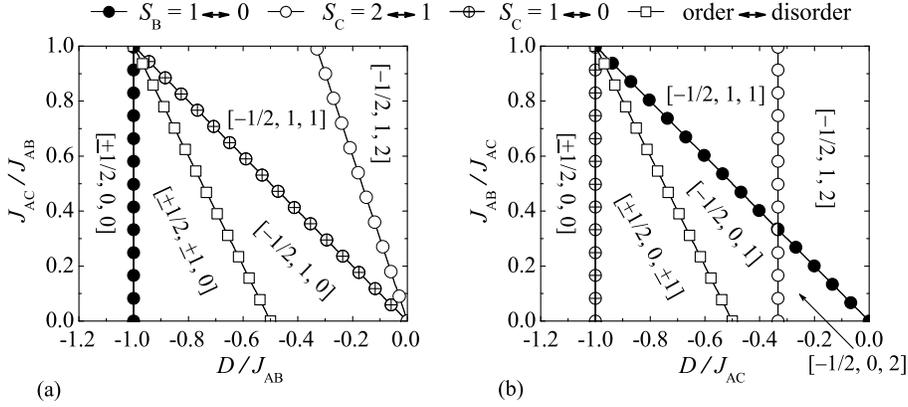,width=1.05\textwidth}}
\vspace*{-0.5cm} \caption{Ground-state phase diagrams in the $D -
J_{\rm AC}$ plane for the system with $J_{\rm AB} \geq J_{\rm AC}$
[Fig.~\ref{fig:2}(a)] and in the $D - J_{\rm AB}$ plane for the
system with $J_{\rm AB} \leq J_{\rm AC}$ [Fig.~\ref{fig:2}(b)]. Spin
order drawn in square brackets indicates typical spin configurations
to emerge within different sectors of the phase diagram. Symbols to
be defined in the legend characterize a spin change that occurs at
the displayed phase boundaries.} \label{fig:2}
\end{figure}

At first, let us take a closer look at a ground-state behaviour of
the decorated model. For this purpose, the ground-state phase
diagrams are illustrated in Fig.~\ref{fig:2} for two particular
cases of the anisotropic spin system. Namely, Fig.~\ref{fig:2}(a)
shows the ground-state phase diagram in the $D - J_{\rm AC}$ plane
for the case when the interaction parameter $J_{\rm AB}$ is stronger
than $J_{\rm AC}$, while Fig.~\ref{fig:2}(b) demonstrates the
ground-state phase diagram in the $D - J_{\rm AB}$ plane for the
case when $J_{\rm AB}$ is weaker than $J_{\rm AC}$. In both these
figures, the spin order drawn in square brackets indicates a typical
spin configuration $[S_{\rm A}, S_{\rm B}, S_{\rm C}]$ to emerge
within the relevant sector of the phase diagram. As one can see, an
eventual spin arrangement is unambiguously determined by a mutual
competition between the exchange interactions $J_{\rm AB}$ and
$J_{\rm AC}$ and the single-ion anisotropy $D$. More specifically,
the interaction parameters $J_{\rm AB}$ and $J_{\rm AC}$
energetically favor the higher spin states of the decorating
spin-$S_{\rm B}$ and spin-$S_{\rm C}$ atoms, while the easy-plane
single-ion anisotropy ($D < 0$) has a tendency to lower their spin
states. Accordingly, when the easy-plane single-ion anisotropy
reinforces, the system exhibits first-order phase transitions
associated with successive lowering of the spin states of decorating
atoms. The general condition allocating the ground-state phase
transition lines accompanied with the spin change $S_{\rm p}
\leftrightarrow S_{\rm p} - 1$ is given by
\begin{equation}
\frac{D_{S_{\rm p}, S_{\rm p} - 1}}{J_{\rm Ap}} = \frac{1}{1 -
2S_{\rm p}}\,,\qquad {\rm p} = {\rm B}, {\rm C}.
\end{equation}

\begin{figure}[bt]
\centerline{\psfig{file=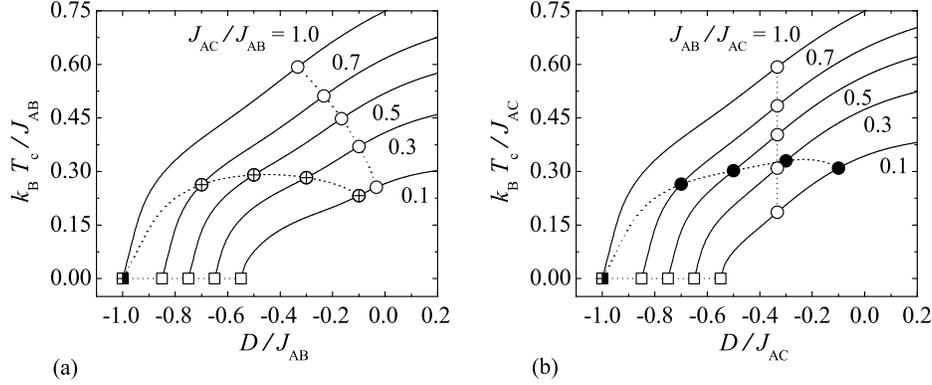,width=1.1\textwidth}}
\vspace*{-0.5cm} \caption{Critical temperature as a function of the
single-ion anisotropy for several values the ratio $J_{\rm
AC}/J_{\rm AB}$ in the system with $J_{\rm AB} \geq J_{\rm AC}$
[Fig.~\ref{fig:3}(a)] and for several values the ratio $J_{\rm
AB}/J_{\rm AC}$ in the system with $J_{\rm AB} \leq J_{\rm AC}$
[Fig.~\ref{fig:3}(b)]. Different symbols characterize the same spin
transition as explained in the legend of Fig.~\ref{fig:2}. Dotted
lines, which connect different spin transitions, are guided for eyes
only.} \label{fig:3}
\end{figure}

Now, we make some comments on the finite-temperature phase diagrams
displayed in Fig.~\ref{fig:3}, which shows the critical temperature
as a function of the single-ion anisotropy for both the investigated
cases. All the phase boundaries depicted as solid lines are the
unique solutions of the critical condition (\ref{eq:Tc}) and in
consequence of that, they represent the lines of the second-order
phase transitions separating the spontaneously ordered phases
(located below the boundary lines) and the disordered paramagnetic
phase stable above the relevant phase boundaries.

As it can be seen from a comparison of Figs.~\ref{fig:2} and
\ref{fig:3}, the quite obvious dependence of the critical
temperature versus single-ion anisotropy arises for the isotropic
case with $J_{\rm AB} = J_{\rm AC}$. Indeed, the critical
temperature monotonically decreases with decreasing the single-ion
anisotropy when $J_{\rm AC}/J_{\rm AB} = 1.0$ (or equivalently
$J_{\rm AB}/J_{\rm AC} = 1.0$) until the spontaneous order
completely vanishes at the boundary value $D_{1,0}/J_{\rm AB} =
-1.0$ ($D_{1,0}/J_{\rm AC} = -1.0$) below which only the disordered
phase may exist. This finding is consistent with our expectation,
since this boundary single-ion anisotropy is strong enough to force
both kinds of decorating spins towards their 'non-magnetic' spin
state $S = 0$.

However, the situation becomes much more striking when assuming
different exchange interactions in the horizontal and vertical
directions ($J_{\rm AC} \neq J_{\rm AB}$). If the interaction
parameter $J_{\rm AB}$ is stronger (weaker) $J_{\rm AC}$, the
critical temperature then monotonically decreases with lowering the
single-ion anisotropy, but it surprisingly does not tend towards the
boundary value $D_{1,0}/J_{\rm AC} = -1.0$ ($D_{1,0}/J_{\rm AB} =
-1.0$) at which all the decorating spin-$S_{\rm C}$ (spin-$S_{\rm
B}$) atoms tend towards their 'non-magnetic' spin state $S = 0$.
Actually, one would intuitively expect that all the phases appearing
in the parameter region $D < - J_{\rm AC}$ ($D < - J_{\rm AB}$)
should be disordered due to their 'quasi-1D' character. The 2D
decorated mixed-spin system indeed behaves as the 'quasi-1D' spin
system when $D < -J_{\rm AC}$ or $D < -J_{\rm AB}$, since it
effectively splits into a set of independent mixed spin-$(S_{\rm A},
S_{\rm B})$ or spin-$(S_{\rm A}, S_{\rm C})$ Ising chains depending
on whether $J_{\rm AB} > J_{\rm AC}$ or $J_{\rm AB} < J_{\rm AC}$ is
considered, respectively.

On the other hand, one should also bear in mind that the system
under investigation is spontaneously long-range ordered if and only
if $\sinh(\beta R_{\rm B}/2)\sinh(\beta R_{\rm C}/2) > 1$, while it
becomes disordered just as $\sinh(\beta R_{\rm B}/2)\sinh(\beta
R_{\rm C}/2) < 1$ in accordance with the critical condition
(\ref{eq:Tc}). In the zero-temperature limit, it can be easily
verified that
\begin{eqnarray}
\lim_{T \rightarrow 0} \Big[ \sinh(\beta R_{\rm B}/2)\sinh(\beta
R_{\rm C }/2) \Big] =
\begin{array}{c}
   \bigg\{
 \end{array}
\begin{array}{cc}
  \infty & \textrm{if}\quad D > - J_{\rm AB}/2 - J_{\rm AC}/2\\
  0 & \,\,\textrm{if}\quad D < - J_{\rm AB}/2 - J_{\rm AC}/2\,,
\end{array}
\end{eqnarray}
what means that the threshold single-ion anisotropy, below which the
system becomes disordered at zero as well as non-zero temperatures,
is given by the following condition $D_{\rm o - d} = - J_{\rm AB}/2
- J_{\rm AC}/2$. This non-trivial phase boundary, which is depicted
in the ground-state phase diagram as a hollow-square line (see
Fig.~\ref{fig:2}), cannot be obtained from simple energetic
arguments. This result directly proves that all the phases that
appear above this order-disorder line are at sufficiently low
temperatures spontaneously ordered in spite of their 'quasi-1D'
nature. In agreement with the aforementioned arguments, the non-zero
critical temperatures to be observed in the parameter space $D < -
J_{\rm AC}$ ($D < - J_{\rm AB}$) manifest the order-disorder
transition between the spontaneously ordered and disordered phases
even though all the decorating spin-$S_{\rm B}$ ($S_{\rm C}$) atoms
reside in the ground state of the ordered phase the 'non-magnetic'
spin state $S = 0$ (see Fig.~\ref{fig:3}). In this respect, the part
of second-order phase transition lines starting at $D = - J_{\rm
AC}$ for $J_{\rm AB} > J_{\rm AC}$ and at $D = - J_{\rm AB}$ for
$J_{\rm AB} < J_{\rm AC}$, and terminating in both the cases at $D =
- J_{\rm AB}/2 - J_{\rm AC}/2$ can be identified as the lines of
critical points at which the spontaneous long-range order of the
'quasi-1D' spin system disappears.

\begin{figure}[bt]
\centerline{\psfig{file=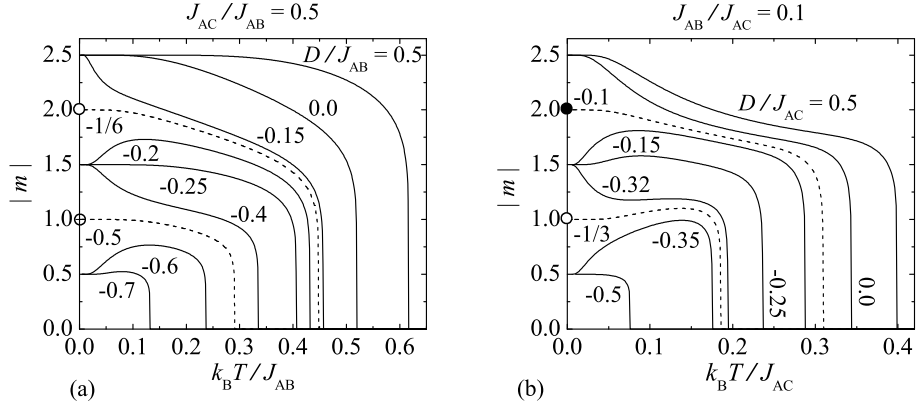,width=1.1\textwidth}}
\vspace*{-0.5cm} \caption{Temperature variation of the total
magnetization $|m|$ for $J_{\rm AC}/J_{\rm AB} = 0.5$
[Fig.~\ref{fig:4}(a)] and $J_{\rm AB}/J_{\rm AC} = 0.1$
[Fig.~\ref{fig:4}(b)]. Different symbols characterize the same spin
transition as explained in the legend of Fig.~\ref{fig:2}. Dotted
lines, which connect different spin transitions, are guided for eyes
only.} \label{fig:4}
\end{figure}

\begin{figure}[bt]
\centerline{\psfig{file=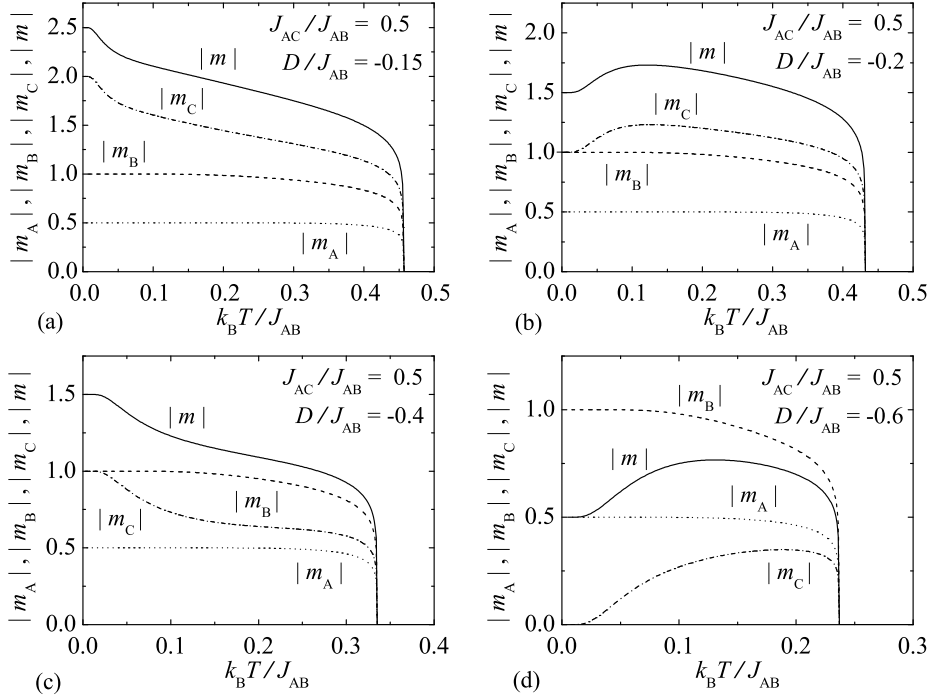,width=1.1\textwidth}}
\vspace*{-0.5cm} \caption{Temperature variation of the sub-lattice
magnetization $|m_{\rm A}|$ (dotted lines), $|m_{\rm B}|$ (dashed
lines), $|m_{\rm C}|$ (dashed-dotted lines) and total magnetization
$|m|$ (solid lines) for the fixed ratio $J_{\rm AC}/J_{\rm AB} =
0.5$ and several values of the the single-ion anisotropy parameter
$D/J_{\rm AB} = -0.15$ [Fig.~5(a)], $-0.2$ [Fig.~5(b)], $-0.4$
[Fig.~5(c)] and $-0.6$ [Fig.~5(d)].} \label{fig:5}
\end{figure}

\begin{figure}[bt]
\centerline{\psfig{file=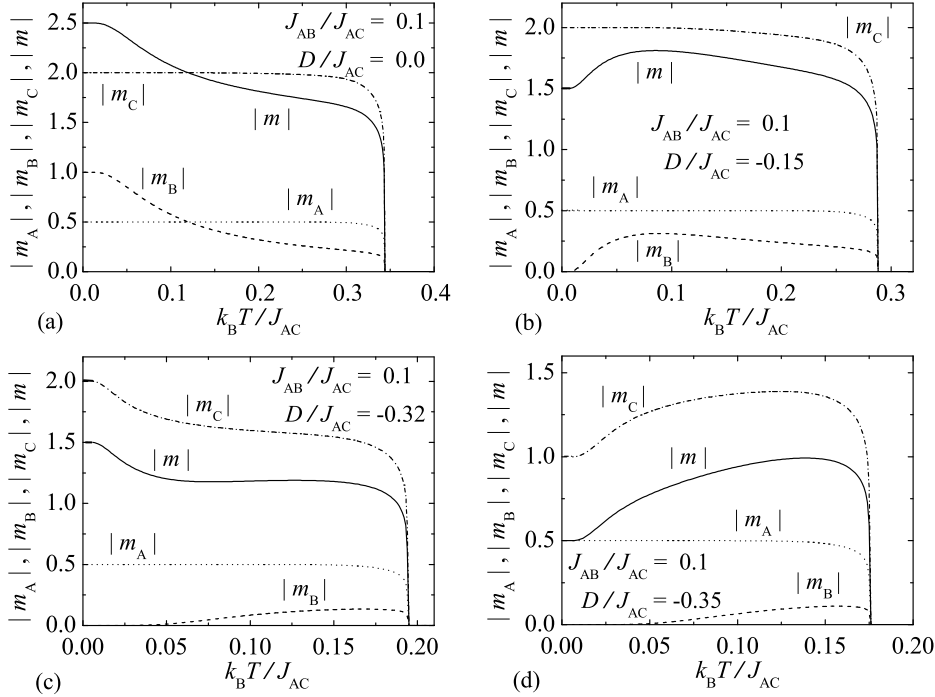,width=1.1\textwidth}}
\vspace*{-0.5cm} \caption{Temperature variation of the sub-lattice
magnetization $|m_{\rm A}|$ (dotted lines), $|m_{\rm B}|$ (dashed
lines), $|m_{\rm C}|$ (dashed-dotted lines) and total magnetization
$|m|$ (solid lines) for the fixed ratio $J_{\rm AB}/J_{\rm AC} =
0.1$ and several values of the the single-ion anisotropy parameter
$D/J_{\rm AC} = 0.0$ [Fig.~6(a)], $-0.15$ [Fig.~6(b)], $-0.32$
[Fig.~6(c)] and $-0.35$ [Fig.~6(d)].} \label{fig:6}
\end{figure}

To provide an independent check of the aforementioned scenario, let
us turn our attention to a thermal variation of the total and
sub-lattice magnetization. The total magnetization $|m| = |m_{\rm A}
+ m_{\rm B} + m_{\rm C}|$ reduced per one site of the original
square lattice is plotted against the temperature in
Fig.~\ref{fig:4} for two investigated particular cases and several
values of the single-ion anisotropy. Note that the curves
corresponding to the coexistence of two different spin orderings (to
emerge in the ground state) are for clarity depicted as dashed
lines. Moreover, it is quite obvious from Fig.~\ref{fig:4} that the
most notable thermal variations of the total magnetization occur in
the vicinity of boundary values of the single-ion anisotropy, which
are associated with the spin change of the decorating atoms in the
ground state. These thermal dependences can be explained with the
help of temperature variations of the sub-lattice magnetization. For
this purpose, all three sub-lattice magnetization are plotted in
Figs.~\ref{fig:5} and \ref{fig:6} against the temperature together
with the total magnetization. For clear presentation, we choose
merely those particular values of the single-ion anisotropy, which
are close enough to the single-ion anisotropies at which the spin
change of decorating atoms occurs when  the same ratio $J_{\rm
AC}/J_{\rm AB}$ and $J_{\rm AB}/J_{\rm AC}$ is fixed as in
Fig.~\ref{fig:4}.

First, let us look more closely on the particular case when $J_{\rm
AB} > J_{\rm AC}$. According to the thermal variations of the
relevant magnetization shown in Fig.~\ref{fig:5}(a), one easily
observes that slightly above the boundary value $D_{2, 1}/J_{\rm AB}
= -1/6$, which corresponds in the ground state to the spin change
$S_{\rm C} = 2 \leftrightarrow 1$, the total magnetization of the
system exhibits a steep initial decrease arising from the thermal
excitations $S_{\rm C} = 2 \rightarrow 1$ of the decorating
spin-$S_{\rm C}$ atoms. As a matter of fact, the rapid temperature
decrease of the sub-lattice magnetization $m_{\rm C}$ can be
detected. By contrast, the opposite temperature-induced excitations
$S_{\rm C} = 1 \rightarrow 2$ are responsible for a rapid increase
of the sub-lattice magnetization $m_{\rm C}$, which in turn causes a
gradual increase in the total magnetization when selecting the
single-ion anisotropy slightly below the aforementioned boundary
value [see Fig.~\ref{fig:5}(b)]. Similar situation can be found in
the vicinity of the boundary value $D_{1, 0}/J_{\rm AB} = -0.5$
associated with the spin change $S_{\rm C} = 1 \leftrightarrow 0$ of
the decorating spin-$S_{\rm C}$ atoms. Actually, the initial
decrease in the total magnetization to be detected at $D/J_{\rm AB}
= -0.4$ [see Fig.~5(c)] can be again attributed to rather rapid
thermal variation of the sub-lattice magnetization $m_{\rm C}$,
which is thermally more easily disturbed than the sub-lattice
magnetization $m_{\rm A}$ and $m_{\rm B}$ due to the low-lying
thermal excitations $S_{\rm C} = 1 \rightarrow 0$. The most
interesting temperature dependence of the total magnetization could
be, however, expected in the parameter space, where the 2D decorated
mixed-spin system effectively behaves as the 'quasi-1D' system due
to its effective splitting into a set of independent mixed
spin-($S_{\rm A}$, $S_{\rm B}$) chains. As it can be seen from
Fig.~\ref{fig:5}(d), the thermal variations of sub-lattice
magnetization really indicate for $D/J_{\rm AB} = -0.6$ the
'quasi-1D' character of the spin system, since this value of the
single-ion anisotropy is strong enough to force all the decorating
spin-$S_{\rm C}$ atoms towards their 'non-magnetic' state $S = 0$ at
the zero temperature. According to this, the sub-lattice
magnetization $m_{\rm C}$ increases from zero with the increase of
temperature on behalf of the predominating temperature-induced
excitations $S_{\rm C} = 0 \rightarrow 1$ of the decorating
spin-$S_{\rm C}$ atoms. Altogether, it might be concluded that a
shape of the total magnetization versus temperature dependence is in
the mixed-spin system with $J_{\rm AB} > J_{\rm AC}$ almost entirely
determined by thermal variations of the sub-lattice magnetization
$m_{\rm C}$.

The situation becomes a little bit more involved when assuming the
system with $J_{\rm AB} < J_{\rm AC}$. A detailed analysis of the
thermal variations of magnetization reveals that the shape of
magnetization curves basically depend on the mutual ratio between
the interaction constants $J_{\rm AB}$ and $J_{\rm AC}$, more
specifically, it depends on whether $J_{\rm AB}/J_{\rm AC} > 1/3$ or
$J_{\rm AB}/J_{\rm AC} < 1/3$. In the former case, the temperature
curvature of the relevant magnetization detected in the vicinity of
boundary single-ion anisotropies are very similar to those formerly
discussed for the system with $J_{\rm AB} > J_{\rm AC}$ and hence,
the discussion concerning with their origin is being omitted here
for brevity. The only difference consists in the fact that the
thermal dependence of sub-lattice magnetization $m_{\rm B}$ now
almost entirely determines the shape of temperature dependence of
the total magnetization in contrast to the case with $J_{\rm AB} >
J_{\rm AC}$, where the sub-lattice magnetization $m_{\rm C}$ has had
a crucial role. However, if one considers $J_{\rm AB}/J_{\rm AC} <
1/3$, the situation becomes even more complicated. The initial
increase (decrease) in the total magnetization found at $D/J_{\rm
AC} = 0.0$ ($-0.15$) can be attributed to the thermal excitations
$S_{\rm B} = 1 \rightarrow 0$ ($S_{\rm B} = 0 \rightarrow 1$) of the
spin-$S_{\rm B}$ atoms. This assertion is strongly supported by the
rapid temperature decrease (increase) of the sub-lattice
magnetization $m_{\rm B}$ as shown in Fig.~\ref{fig:6}(a)
[Fig.~\ref{fig:6}(b)]. Moreover, according to the magnetization
curves depicted in Fig.~\ref{fig:6}(b) one easily finds that the
value of the single-ion anisotropy $D/J_{\rm AC} = -0.15$
corresponds to the ground-state region, where all the decorating
spin-$S_{\rm B}$ atoms are driven towards their 'non-magnetic' state
$S = 0$. In consequence of that, the 2D mixed-spin system should
effectively behave as a set of independent mixed spin-($S_{\rm A}$,
$S_{\rm C}$) chains (as the 'quasi-1D' system) having the decorating
spin-$S_{\rm C}$ atoms in the 'magnetic' spin state $S = 2$.
Actually, it is easy to observe from Fig.~Fig.~\ref{fig:6}(b) that
the sub-lattice magnetization $m_{\rm B}$ indeed starts from zero in
agreement with our expectations and it is merely the effect of
temperature that causes its initial rise. Another notable thermal
dependence of the total magnetization can be observed at $D/J_{\rm
AC} = -0.32$, which is slightly above the boundary value $D_{2,
1}/J_{\rm AC} = -1/3$ associated with the spin change $S_{\rm C} = 2
\leftrightarrow 1$. As one can see from Fig.~\ref{fig:6}(c), the
total magnetization firstly rapidly decreases and only then slightly
increases upon further temperature increase. This interesting
non-monotonous behaviour of the total magnetization can be explained
as a result of the initial thermal excitations $S_{\rm C} = 2
\rightarrow 1$ of the decorating spin-$S_{\rm C}$ atoms, while the
temperature-induced excitations $S_{\rm B} = 0 \rightarrow 1$ of the
decorating spin-$S_{\rm B}$ atoms overwhelm in the region of higher
temperatures. It is indeed quite apparent from Fig.~\ref{fig:6}(c)
that the sub-lattice magnetization $m_{\rm B}$ increases much more
rapidly in this temperature region than the sub-lattice
magnetization $m_{\rm C}$ declines. This behaviour consequently
leads to a slight increase in the total magnetization. Finally, the
total magnetization of the system gradually increases with the
temperature even if $D/J_{\rm AB} = -0.35$ is fixed. In this
parameter region, the 2D decorated mixed-spin system again
effectively behaves as the 'quasi-1D' system due to its effective
splitting into a set of independent mixed spin-($S_{\rm A}$, $S_{\rm
C}$) chains, but now with the decorating spin-$S_{\rm C}$ atoms in
the 'magnetic' state $S = 1$. The observed thermal increase in the
total magnetization can be obviously attributed to a rapid thermal
variation of the sub-lattice magnetization $m_{\rm B}$ and $m_{\rm
C}$ arising from the temperature-induced spin excitations $S_{\rm B}
= 0 \rightarrow 1$ and $S_{\rm C} = 1 \rightarrow 2$ of the
decorating spin-$S_{\rm B}$ and spin-$S_{\rm C}$ atoms,
respectively.

\section{Conclusion}
\label{conclusion} In the present article, the mixed-spin Ising
model on a decorated square lattice with two different decorating
spins of integer magnitudes $S_{\rm B}$ and $S_{\rm C}$ ($S_{\rm B}
\neq S_{\rm C}$) placed on horizontal and vertical lattice bonds,
respectively, has been investigated within the framework of the
generalized decoration-iteration transformation. By the use of this
exact mapping procedure, the exact solution for the system under
investigation has been attained by establishing a simple mapping
relationship with the spin-$1/2$ Ising model on the corresponding
anisotropic square (rectangular) lattice, whose exact solution is
known since Onsager's pioneering work\cite{Ons44}. Besides, the
aforedescribed mapping transformation method enjoys also immense
practical importance, because it is rather general and it implies a
possibility of further interesting extensions. Actually, this
analytical approach can be straightforwardly extended to account
also for: (i) the interaction between next-nearest-neighbouring
spins $S_{\rm A}$; (ii) the multi-spin interaction between the
decorating spins and their nearest neighbours; (iii) the biaxial
single-ion anisotropy acting on the decorating atoms; (iv) other
decorated lattices, such as decorated honeycomb or triangular
lattices with two or three distinct decorating spins; (v) decorated
lattices with two or more decorating spins per one lattice bond;
etc. Furthermore, it is also noteworthy that the procedure can be in
principle applied to 3D decorated models as well. Nevertheless, we
cannot present exact results for any 3D decorated models, since the
exact solution of the spin-$1/2$ Ising model on the corresponding 3D
lattice is not known so far.

The most interesting result to emerge from the present study
consists in providing an exact evidence for the spontaneous
long-range order, which surprisingly appears in the 2D decorated
spin system in spite of its 'quasi-1D' character. As a matter of
fact, we have found that the 2D decorated system remains
spontaneously ordered even if one sub-lattice of decorating spins
(either $S_{\rm B}$ or $S_{\rm C}$) is forced by a sufficiently
strong single-ion anisotropy to be 'non-magnetic' and the system
becomes 'quasi-1D' due to the effective splitting into a set of
independent mixed spin-($S_{\rm A}$, $S_{\rm B}$) or spin-($S_{\rm
A}$, $S_{\rm C}$) chains depending on $J_{\rm AB}$ is stronger or
weaker than $J_{\rm AC}$, respectively. This finding has an obvious
relevance to the understanding of the 'quasi-1D' spin systems
inclinable to the spontaneous long-range ordering below some
critical temperature, which necessarily do not need to arise from
interactions establishing 2D or 3D magnetic structure, but may
represent an inherent feature of the 'quasi-1D' system. From this
point of view, the exactly solved mixed-spin system presented in
this article would be important if some experimental realization of
it would confirm the spontaneous ordering notwithstanding of its
'quasi-1D' character.

Although the magnetic structure description of the present
mixed-spin Ising model may not be fully realistic for true magnetic
materials, it is quite reasonable to expect that the exact solution
of this simplified model illustrates many important vestiges of the
real critical behaviour. Moreover, the exact solution of theoretical
model system may also bring other valuable insights into the
thermodynamical properties (magnetization, entropy, specific heat)
of true magnetic materials without applying any crude and/or
uncontrollable approximative theories.

According to this, the main stimulus for the study of the mixed-spin
Ising model on the square lattice with two different decorating
spins of magnitudes $S_{\rm B}$ and $S_{\rm C}$ placed on horizontal
and vertical bonds, respectively, can be viewed in connection with
its possible experimental realization. It is therefore of particular
interest to mention that polymeric compounds with the architecture
of decorated square network have been rather frequently prepared in
an attempt to design novel bimetallic coordination compounds.
Indeed, the magnetic structure of the decorated square lattice can
be found in two numerous series of polymeric coordination compounds
with the following general formula: $[{\rm Ni}({\rm L})_2]_2 [{\rm
Fe}({\rm CN})_6]{\rm X}.{\rm nH}_2{\rm O}$\cite{Ohb95} and ${\rm
A}[{\rm M_B}({\rm L})]_2 [{\rm M_A}({\rm CN})_6].{\rm nH}_2{\rm O}$
(${\rm M_A} = {\rm Fe}$, ${\rm Mn}$, ${\rm Cr}, {\rm Co}$; ${\rm
M_B} = {\rm Fe}$, ${\rm Mn}$)\cite{Miy95}. In the former series, the
magnetic ${\rm Fe}^{3+}$ ($S_{\rm A} = 1/2$) ions reside the square
lattice sites and ${\rm Ni}^{2+}$ ($S_{\rm B} = 1$) ions decorate
each its bond, while in the latter series the low-spin ${\rm
M_A}^{3+}$ ions such as ${\rm Fe}^{3+}$ ($S_{\rm A} = 1/2$), ${\rm
Mn}^{3+}$ ($S_{\rm A} = 1$), or ${\rm Cr}^{3+}$ ($S_{\rm A} = 3/2$)
reside the square lattice sites and the high-spin ${\rm M_B}^{3+}$
ions like ${\rm Mn}^{3+}$ ($S_{\rm B} = 2$) or ${\rm Fe}^{3+}$
($S_{\rm B} = 5/2$) occupy the decorating lattice sites. Up to now,
we are not aware of any trimetallic polymeric compound whose network
assembly would consist of two different kinds of magnetic metal ions
(decorating spins $S_{\rm B} \neq S_{\rm C}$) placed on the square
net made up by the third magnetic metal ion (by the spin $S_{\rm
A}$), but the vast number of the bimetallic coordination compounds
from the aforementioned series gives us hope that a targeted
synthesis of such trimetallic compounds could be successfully
accomplished in the near feature.

\section*{Acknowledgements}

The authors would like to express their gratitude to scientific
grant agencies for financial support given under the grants VVGS
12/2006, VEGA 1/2009/05 and APVT 20-005204.

\end{document}